# THE AGE OF THE UNIVERSE[1]


Sidney van den Bergh

Dominion Astrophysical Observatory

5071 West Saanich Road

Victoria, British Columbia

V8X 4M6, Canada


## ABSTRACT


It is found that the ages of the most metal-poor Galactic globular clusters are probably $\approx$ 18 Gyr. On the other hand $H_o = 82 \pm 8$ km s$^{-1}$ Mpc$^{-1}$ (in conjunction with the standard Einstein - de Sitter model with $\Omega = 1$ and $\Lambda = 0$) yields an age of the Universe $t_o = (\tfrac{2}{3})\ H_o^{-1} \approx 8$ Gyr. Some possible reasons for this discrepancy are explored.

## 1.    INTRODUCTION

The age of the Universe can be determined from extragalactic observations which yield both the Hubble parameter $H_o$ and $f$ in the equation

$$t_o \, H_o = f \, (q_o, \, \lambda_o). \tag{1}$$

A less ambitious approach is to derive a minimum age of the Universe from the ages of the most ancient individual stars and clusters in the Galaxy.  Due to possible second parameter effects, the most reliable ages of old Galactic globular clusters are probably derived by using calibrations based on Galactic RR Lyrae variables that have halo kinematics.  Only weak constraints on the age and evolution of the Universe are provided by radioactive age dating of the elements, which is sometimes referred to as cosmochronology.  Possible problems with the Cepheid Period-Luminosity relation and with distances of supernovae of type Ia will also be discussed.

## 2.    DISTANCE OF THE LMC

The distance to the Large Magellanic Cloud affects estimates of the age of the Universe in two ways.  In the first place, values of $H_o$ (and, hence, the age of the Universe $t_o$ ) determined via observations of Cepheids in the Virgo Cluster (Pierce et al. 1994, Freedman et al. 1994) were calibrated using an LMC distance of 50 Mpc, corresponding to a true distance modulus $(m\text{-}M)_o = 18.49$.  An increase in the distance to the Large Cloud would decrease $H_o$ and therefore increase $t_o$ .  Secondly, the distance to the LMC might be used to set the luminosities of RR Lyrae stars, which in turn determine the evolutionary ages calculated for globular clusters. Unfortunately, the distance to the Large Cloud remains controversial.  From multi-color photometry of Cepheids, Laney & Stobie (1992) find a true LMC distance modulus $(m\text{-}M)_o = 18.57 \pm 0.04$.  This is consistent with work by Gieren, Barnes & Moffet (1993) and by Gieren, Richtler & Hilker (1994) who use the visual surface brightness modification of the Baade-Wesselink method to derive $(m\text{-}M)_o = 18.6 \pm 0.1$ and $18.47 \pm 0.20$, respectively.



However, a much smaller distance modulus $(m-M)_o = 18.23 \pm 0.04$ is obtained if the Galactic calibration of the luminosities of halo RR Lyrae stars is applied to the cluster-type variables (Walker 1992) in the Large Cloud.

Could this discrepancy be due to the effects of metallicity on the luminosities of Cepheids? In other words, could the Galactic Period-Luminosity relation be giving the wrong distance for Cepheids in the Large Cloud? Assuming [Fe/H] = -0.3 (Z = 0.009) for young stars in the LMC (Spite & Spite 1991), in conjunction with $\triangle Y / \triangle Z = 3$, and Y (primordial) = 0.228 (Pagel 1992), the classical evolutionary models of Chiosi et al. (1993) yield a Period-Luminosity relation for the Large Cloud that is only 0.01 mag brighter for 10-day Cepheids than it is for similar objects in the Galaxy. This indicates that there is presently no reason to suspect that the Cepheid Period-Luminosity relation for the LMC gives a distance modulus that is seriously in error. Apparent confirmation of this conclusion is provided by Gould (1995) who used observations of the ring surrounding SN 1987A to obtain a geometrical distance D = $47.3 \pm 0.8$ kpc (circular ring assumed), corresponding to $(m-M)_o = 18.37 \pm 0.04$. This value is in reasonable agreement with the Cepheid results obtained by Laney & Stobie (1992) who found $(m-M)_o = 18.57 \pm 0.04$. The combined Cepheid and SN 1987A data therefore suggest $(m-M)_o \approx 18.5$ for the Large Cloud, which is strongly discordant with the value $(m-M)_o = 18.23 \pm 0.04$, that was obtained by Walker (1992) from the Galactic calibration of halo RR Lyrae stars applied to Walker's (1992) observations of the Cluster-type variables in the LMC.

Hydrodynamic pulsation models for three RRc stars in the LMC (Simon & Clement 1993), when matched to the pulsation period P and Fourier phase parameter $\varphi_{31}$, yield



$(m-M)_o = 18.54$. This value is in good agreement with the Large Cloud distance modulus derived from Cepheids and in reasonable agreement with that derived from SN 1987A.

How is one to account for the discordant distance obtained by Walker (1992)? Possibly, the RR Lyrae stars in the Galactic halo and in the Large Cloud obey different $M_V$ (RR) versus [Fe/H] relations. It is well known (van den Bergh 1965, Sandage & Wildey 1967) that the population gradients along the horizontal branches of globular clusters are a function of both [Fe/H] and of a second parameter that has not, as yet, been identified. By the same token, the luminosities of RR Lyrae stars might be a function of both [Fe/H] and one or more additional parameters. It is tentatively suggested that such a second parameter effect might account for the differences between the Cepheid and RR Lyrae distance moduli for some (but not all!) Local Group members that have been discussed in detail by van den Bergh (1995). In view of the fact that the $M_V$ versus [Fe/H] relations might differ in the LMC and in the Galaxy, it seems safest to use the $M_V$ (RR) calibration derived from Galactic halo RR Lyrae stars to calculate the ages of Galactic halo globular clusters.

## 3.    AGES OF GLOBULAR CLUSTERS

The ages calculated for globular clusters depend on both the input physics and on the adopted magnitudes for the cluster horizontal branches i.e. $M_V$(RR). Recently, Chaboyer (1995) has discussed the uncertainties in cluster ages that result from reasonable changes in (1) the mixing length, (2) color transformations, (3) model atmospheres, (4) opacities, (5) convective overshooting, (6) nuclear reaction rates, (7) helium diffusion, (8) uncertainty in the helium abundance, (9) errors in [Fe/H], and (10) changes in [$\alpha$/Fe]. He finds that such factors result in a combined uncertainty of only $\pm$ 15% (m.e.) in the calculated cluster ages. However, larger errors result from the uncertainty in cluster horizontal branch magnitude which can change the calculated ages from 18.2 $\pm$ 0.6 Gyr for the faint RR Lyrae calibration adopted by



Layden, Hawley & Hanson (1995) to $14.2 \pm 0.5$ Gyr for the bright horizontal branch calibration of Walker (1992). <u>The ~ 25% age uncertainty resulting from possible errors in the magnitude level of globular cluster horizontal branches is therefore more significant than errors that might result from changes to the input physics.</u>

From a comprehensive study of Galactic RR Lyrae stars using the Baade-Wesselink technique Carney, Storm & Jones (1992) find that

$$M_V (RR) \; = \; (0.15 \pm 0.01) \; [Fe/H] + 1.01 \pm 0.08. \tag{2}$$

A much steeper dependence of $M_V$ (RR) on [Fe/H] had previously been found by Sandage & Cacciari (1990) who obtained

$$M_V (RR) \; = \; 0.39 \; [Fe/H] + 1.27. \tag{3}$$

The best way to derive both the zero-point and the slope of this relation is via HST observations of the horizontal branches of globular clusters of known metallicity in the halo of M 31, which are all located at approximately the same distance. To date, only two globular clusters in the Andromeda nebula have been studied in this way. From the metal-rich cluster K 58 ([Fe/H] = -0.57) and the metal-poor cluster K 219 ([Fe/H] = -1.83) observed by Ajhar et al. (1994) one obtains d $M_V$ (RR) / d [Fe/H] = +0.12, which clearly favors the relation given in Eqn. (2) over that in Eqn. (3). It is noted in passing that a large value of d $M_V$ (RR) / d [Fe/H] would reduce the calculated ages of old metal-poor clusters (which have luminous horizontal branches), and increase the ages of younger metal-rich globulars (which have faint horizontal branches). The net effect of a large value of d $M_V$ (RR) / d [Fe/H] is therefore that it decreases the calculated age spread among globular clusters. This accounts for the small age range of Galactic globular clusters found by Mazzitelli, D'Antona & Caloi (1995).



Force-fitting their data to the slope of Eqn. (2), Layden, Hawley & Hanson (1995) find

$$M_V \text{ (RR)} = 0.15 \text{ [Fe/H]} + 0.99 \pm 0.12 \tag{4}$$

from the statistical parallaxes of RR Lyrae stars. Note, in particular, that the zero-point of Eqn. (4) is in excellent agreement with that which Carney et al. (1992) obtained using the Baade-Wesselink method. <u>The fact that these two independent techniques yield essentially the same zero-point strengthens confidence in the adopted relationship between $M_V$ and [Fe/H] for Galactic RR Lyrae variables</u>.

It is therefore concluded that presently available data favor a relation of the form

$$M_V \text{ (RR)} \approx 0.15 \text{ [Fe/H]} + 1.0 \tag{5}$$

for Galactic halo RR Lyrae variables. With this calibration (Chaboyer 1995) the ages of the oldest Galactic globular clusters are $\approx$ 18 Gyr.

## 4.    AGE OF THE GALACTIC DISK

Giant molecular clouds eventually destroy most open clusters in the Galactic disk. As a result, the number of clusters that survive for a long period of time is expected to be quite small. Among the oldest known disk clusters are NGC 6791 and NGC 188. For NGC 188, Dinescu et al. (1995) derive an age of $6.0^{+1.0}_{-0.5}$ Gyr. From a comparison of color-magnitude diagrams, Kaluzny & Rucinski (1995) find that NGC 6791 is 1 Gyr older than NGC 188. On the other hand, Tripicco et al. (1995) find an age of $10.0 \pm 0.5$ Gyr for this cluster. The age of the oldest known open cluster is therefore 7 - 10 Gyr. Since even older open clusters might exist, this value represents a lower limit to the age of the Galactic disk.

White dwarfs that formed from the first generation of Galactic disk stars have not yet had time to cool and fade into invisibility. As a result, the white dwarf luminosity function has written into it the age and star formation history of the Galactic disk (Wood 1992). Using the



best current inputs, Wood derives a disk age of 7.5 Gyr to 11 Gyr.  Xu and Van Horn (1992)

find that taking the energy release associated with O/C, Fe/C and Ne/C phase transitions in

predominantly C white dwarfs into account will increase the maximum age of the Galactic disk

near the sun to ~ 13 Gyr.  On the other hand, models for white dwarfs that are covered by a

thick layer of hydrogen are found (Wood - private communication) to be 1 Gyr to 1.5 Gyr

younger than models covered by only a thin (or no) hydrogen layer.  Unfortunately, some doubt

is cast on all of these white dwarf cooling time calculations by the observation (von Hippel,

Gilmore & Jones 1995) that white dwarfs in some open clusters have cooling ages that are not

consistent with the isochrone ages of their parent clusters.  In any case, it is encouraging that <u>all</u>

<u>of these Galactic disk ages are significantly smaller than the age of about</u> 18 <u>Gyr obtained in</u> § 3

<u>for the oldest and most metal deficient halo globular clusters</u>.

## 5.  COSMOCHRONOLOGY

Nuclear cosmochronology uses measurements of the ratios of long-lived isotopes such as

$^{187}$Re, $^{232}$Th, $^{238}$U and $^{235}$U, in conjunction with calculations of the  initial production ratios of

such elements, to calculate the age of the elements in the Galaxy.  In general, such calculations

require models for the evolution of the rate of star formation in addition to the nuclear input

parameters.  A very detailed review of these problems has been given by Cowan, Thielemann &

Truran (1991) who conclude that presently available observational constraints favor a Galactic

age in the range of 10 - 20 Gyr.

## 6.  THE HUBBLE PARAMETER

The Hubble parameter $H_o$ has dimension $t^{-1}$ and therefore provides important constraints

on the age of the Universe.  In a recent review (van den Bergh 1994) of all data available prior

to 1994.5, it was found that most high-weight distance indicators suggested that



$H_o \gtrsim 75$ km s$^{-1}$ Mpc$^{-1}$. This conclusion has subsequently been greatly strengthened by observations of Cepheid variables in galaxies associated with the Virgo cluster. First, Pierce et al. (1994) observed three Cepheids in NGC 4571 with the Canada-France-Hawaii Telescope. Subsequently, Freedman et al. (1994) used the Hubble Space Telescope to obtain more detailed observations of 12 Cepheids on M 100. Both groups of observers used the same Cepheid Period-Luminosity relation and assumed a distance modulus (m-M)$_o$ = 18.5 for the Large Magellanic Cloud. From their HST observations, Freedman et al. (1994) obtained a distance of 17.1 Mpc for M 100, whereas Pierce et al. (1994) derived D = 14.9 Mpc for the distance to NGC 4571. Estimates of the error budgets in the distance modulus derived from R-band observations of NGC 4571 by Pierce et al. and form I-band observations of M 100 by Freedman et al. are compared in Table 1. Not unexpectedly, the HST observations yield a distance modulus (m-M)$_o$ = 31.16 ± 0.15 for M 100 that is more precise than the value (m-M)$_o$ = 30.87 ± 0.18 that Pierce et al. obtained for NGC 4571. Nevertheless, the observations of NGC 4571 may give a more accurate determination of the true distance to the center of the Virgo cluster because this galaxy, which is situated only 2°.7 from M 87, exhibits significant HI stipping and is therefore likely to be physically associated with the Virgo Core. On the other hand, the gas-rich spiral M 100 might be located in the outer envelope of the Virgo cluster. If one (more-or-less arbitrarily) assumes that NGC 4571 and M 100 are within 0.5 Mpc and 1.5 Mpc of the core of Virgo, respectively, then the total errors of the Virgo distance moduli become (m-M)$_o$ = 30.87 ± 0.19 from NGC 4571, and (m-M)$_o$ = 31.16 ± 0.24 from M 100. In light of these results, it will be assumed that (m-M)$_o$ = 31.0 ± 0.2 for the Virgo cluster.

The magnitude of the retardation of the Local Group by the mass of the Virgo cluster is not well determined. It therefore appears most prudent to anchor the determination of the Hubble parameter to the more distant Coma cluster. The true difference △ (m-M) between the



distance moduli of the Coma and Virgo clusters is quite well determined (van den Bergh 1992). With a Virgo modulus $(m-M)_o = 31.0 \pm 0.2$ and $\triangle (m-M) = 3.71 \pm 0.05$, one obtains a true Coma modulus $(m-M)_o = 34.71 \pm 0.21$, corresponding to a distance of $87 \pm 8$ Mpc. In conjunction with a Coma velocity (corrected to the microwave rest frame) of $7146 \pm 86$ km s$^{-1}$, this yields $H_o = 82 \pm 8$ km s$^{-1}$ Mpc$^{-1}$. For the standard Einstein - de Sitter $\Omega = 1$, $\Lambda = 0$ cosmology one then obtains an age $t_o = (\frac{2}{3})$ $H_o^{-1} = 6.52$ h$^{-1}$ Gyr $= 8.0$ Gyr. It is by now well known that such a low value is strongly contradicted by the ages of the oldest globular clusters and weakly contradicted by cosmochronology.

## 7.   GRAVITATIONAL LENSES

The time delay between brightness variations of the components of a gravitational lens may be used, in conjunction with a model for the mass distribution surrounding the lensing object, to derive a value for the Hubble parameter. From a time delay of $423 \pm 6$ days between the components of the quasar $0957 + 561$, Pelt et al. (1995) find that $14 < H_o$ (km s$^{-1}$ Mpc$^{-1}$) $< 67$.

## 8.   SUPERNOVAE

### a.  Supernovae of Type II

Schmidt et al. (1994) have obtained detailed photometric and spectroscopic observations of the distant type II supernova 1992am. Application of the expanding photosphere method to this object yields a distance D $= 180\,^{+30}_{-25}$ Mpc. In conjunction with its radial velocity of 14600 km s$^{-1}$, this yields a Hubble parameter $H_o = 81\,^{+17}_{-15}$ km s$^{-1}$ Mpc$^{-1}$. Determinations of $H_o$ using less distant SNe II might be affected by uncertainties resulting from deviations of their parent galaxies from a smooth Hubble flow. Photometric and spectroscopic observations of additional very distant SNe II would clearly be of great value.



**b.  Supernovae of Type Ia**

In recent years, a number of authors have used SNe Ia in attempts to calibrate the extragalactic distance scale.  Such distance determinations are ultimately based on (1) the luminosity calibration of a few relatively nearby SNe Ia, and (2) the assumption that supernovae of type Ia are good "standard candles"; or at least on the hypothesis that a sub-set of SNe Ia can be found which are standard candles.

Because supernovae of type Ia are so luminous, they would, if they were good standard candles, be almost ideal distance indicators.  However, observations collected during the last decade clearly show that SNe Ia are <u>not</u> all identical and that some of them have luminosities that deviate significantly from the mean.  For example, SN 1957B and SN 1991bg, which both occurred in the (dust free) elliptical galaxy M 84, had luminosities at maximum light that differed by $\triangle$ B (max) $\approx$ 2.4 mag.  Hamuy et al. (1994) show that ¼ to ½ of all SNe Ia <u>per unit volume</u> may be subluminous objects resembling SN 1991bg and SN 1992K.  However, such intrinsically faint supernovae will be under-represented in magnitude-limited supernova searches.

Attempts to use SNe Ia as standard candles have followed two alternate paths.  Branch, Fisher & Nugent (1993) have used photometric and spectroscope data in an attempt to weed out subluminous and (or) peculiar SNe Ia.  However, the work done by Maza et al. (1994) shows that even apparently normal supernovae can differ in luminosity by as much as 0.8 mag.  Alternatively, Phillips (1993) has attempted to fit all SNe Ia to a single maximum magnitude versus rate of decline relation.  The usefulness of such a relation depends on the residual intrinsic dispersion of individual SNe Ia about the adopted maximum magnitude versus rate-of-decline relationship.  The fact that SN 1885 (S And), which was probably an SN Ia, deviates significantly from the relation adopted by Phillips (1993) is clearly a source of concern.  Riess,



Press & Kirshner (1995) find that they can reduce the dispersion in the distances calculated for SNe Ia to 0.2 mag by considering not just the rate of decline, but the entire shape of the supernova light curve within three weeks of maximum light. It will be interesting to see how this approach holds up once a larger set of supernova observations becomes available.

Recently, Höflich & Khokholov (1995) have calculated theoretical models for SNe Ia produced by deflagrations, detonations, delayed detonations, pulsating delayed detonations and tamped detonations of Chandrasekhar mass progenitors. By fitting individual well-observed supernovae (for which reddening values can be estimated) to their models, Höflich & Khokholov find $H_o = 67 \pm 9$ km s$^{-1}$ Mpc$^{-1}$.

Fig. 1  $M_V$ versus B-V at maximum light for models of supernovae of type Ia calculated by Höflich & Khokholov (1995). The arrow has the slope of a reddening line with $A_V = 3.1$ $E_{B-V}$.

A different approach is suggested by Fig 1, which shows $M_V$(max) versus (B-V) at maximum light for 36 models calculated by Höflich & Khokholov (1995). This Figure shows that the faintest SNe Ia are also the reddest, which is in agreement with observations which show that the faint supernovae 1991bg and 1992K were intrinsically red. By coincidence, the slope of the $M_V$ versus B-V relation for SN Ia models is very similar to that produced by interstellar reddening, for which $A_V = 3.1$ $E_{B-V}$. It is therefore possible to define a parameter



$$M_V^\bullet = M_V - 3.1\ (B-V)\ , \tag{6}$$

which is almost independent of both reddening <u>and</u> model-dependent SN Ia luminosity differences. For 36 supernova models by Höflich & Khokholov $< M_V^\bullet > = -19.60 \pm 0.05$, with a dispersion $\sigma = 0.29$ mag. For 13 SNe Ia observed by Hamuy et al. (1995) $< M_V^\bullet > + 5 \log (H_o /85) = -18.83 \pm 0.11$, with a dispersion $\sigma = 0.40$ mag. Combining these two results yields $H_o = 59.6 \pm 3.3$. Note that this value of the Hubble parameter depends only on the input physics used by Höflich & Khokholov, and that it is therefore independent of any astronomical calibrations.

The Hubble Space Telescope has so far been used to determine the distances of IC 4182 and of NGC 5253. Using the Cepheid Period-Luminosity relation, Saha et al. (1994) obtain a distance modulus $(m-M)_o = 28.36 \pm 0.09$ for IC 4182, which produced the type Ia supernova 1937C. Furthermore, Saha et al. (1995) find $(m-M)_V = 28.10 \pm 0.10$ for NGC 5253, which has produced SN 1895B and SN 1972E; both of which were of type Ia. Unfortunately, these new results have not yet resolved the controversy which presently surrounds the luminosity calibration of SNe Ia. Saha et al. (1994) find $H_o = 52 \pm 9$ km s$^{-1}$ Mpc$^{-1}$ by comparing SN 1937C with distant SNe Ia. However, Pierce & Jacoby (1994) derive $H_o = 74 \pm 6$ km s$^{-1}$ Mpc$^{-1}$ by re-measuring the original photographic plates of SN 1937C, and by applying the Phillips' (1993) maximum magnitude versus rate of decline relation to its light curve. From re-measurement of the original plates of SN 1985B in NGC 5253, Schaefer (1995) finds $26 \lesssim H_o$ (km s$^{-1}$ Mpc$^{-1}$) $\lesssim 61$. From V-band observations of SN 1972E in NGC 5253 (but without application of a rate-of-decline correction), Saha et al. (1995) derive $H_o \simeq 57$ km s$^{-1}$ Mpc$^{-1}$. Taken at face value, these results suggest that SNe Ia yield a lower value of the Hubble parameter than do most other techniques.



Alternatively, one might assume that the SNe Ia in IC 4182 and NGC 5253 were of above-average luminosity at maximum light. Although ad hoc this suggestion is not entirely unreasonable. Branch & van den Bergh (1993) found that the expansion velocities of SNe Ia in galaxies that contain a young stellar population tend to be higher than those of SNe Ia that occur in galaxies in which the dominant stellar population is old. Furthermore, Hamuy et al. (1995) find that "galaxies having a younger population appear to host the most luminous events corresponding to the most massive progenitors". Such a bias would lower the value of $H_o$ derived from IC 4182 and NGC 5253. An additional source of uncertainty is that neither the slope, nor the intrinsic dispersion, of the maximum magnitude versus rate of decline relation corrections (which need to be applied to SN 1895B, SN 1937C and SN 1972E) are well determined.

## 9.     AGES OF ELLIPTICAL GALAXIES

Recently, Oke, Gunn & Hoessel (1995) have observed the spectral energy distribution of the nearby giant elliptical NGC 4889. From a comparison of their observations with the evolutionary tracks of Bruzual & Charlot (1993), these authors conclude that the dominant stellar population in this object has an age of $14 \pm 2$ Gyr. Allowing a period of $\sim 1$ Gyr for assembly of giant ellipticals and the formation of the bulk of their stars, this gives an age of the Universe to $\approx 15 \pm 2$ Gyr. It is of particular interest to note that Oke et al. find a much younger age of 4 - 8 Gyr for the giant elliptical galaxies in eight distant clusters at a redshift $z = 0.5$ Unfortunately, such age determinations of distant galaxies are not yet sufficiently accurate to place meaningful constraints on cosmological models.

## 10.     SUMMARY AND CONCLUSIONS

The Large Magellanic Cloud is perhaps the most crucial stepping stone along the difficult path that ultimately leads to the calibration of the extragalactic distance scale. It is therefore



disconcerting that the three most highly weighted distance determinations for the LMC give
discordant results:

| Cepheids | Laney & Stobie (1992) | $(m-M)_o = 18.57 \pm 0.04$ |
| SN 1987A ring | Gould (1995) | $(m-M)_o = 18.37 \pm 0.04$ |
| $M_V$ (RR) | Walker (1992) | $(m-M)_o = 18.23 \pm 0.04$ |

It would clearly be very important to (1) improve the Galactic calibration of the Cepheid Period-
Luminosity relation, and (2) strengthen and confirm the work by Chiosi et al. (1993), which
appears to show that differences in Y and Z are not likely to produce significant differences
between the Galactic and LMC Period-Luminosity relation.

The discrepant distance derived from application of the Galactic $M_V$ (RR) calibration to
the cluster-type variables in the Large Cloud might be understood if $M_V$ (RR) is a function of both
[Fe/H] and a second parameter.  It seems safest to use the calibration of RR Lyrae stars of the
Galactic halo population to derive $M_V$ (RR) in Galactic globular clusters.  Doing this yields an age
of ~ 18 Gyr for the oldest and most metal-deficient globular clusters.  Chaboyer (1995) has
shown that reasonable changes in the mixing length, color transformations, model atmospheres,
opacities, convective overshooting, nuclear reaction rates, helium diffusion, uncertainty in helium
abundance, errors in [Fe/H] and changes in [$\alpha$/Fe] will change this age by less than one Giga year.
This value is consistent with both cosmochronology and with white dwarf ages in the Galactic
disk.

From two independent observations (Pierce et al. 1994, Freedman et al. 1994) of
Cepheids in the Virgo cluster, $H_o = 82 \pm 8$ km s$^{-1}$ Mpc$^{-1}$.  For a Universe with $\Omega = 1$ and $\Lambda = 0$,
this yields an age $t_o = \frac{2}{3}$ $(H_o^{-1}) = 8.0$ Gyr, which clearly conflicts with the ages of the oldest
globular clusters.  The most distant supernova of type II (Schmidt et al. 1994) yields



$H_o = 81 \pm 16$ km s$^{-1}$ Mpc$^{-1}$, which agrees well with the Cepheid determination of $H_o$. However, supernovae of type Ia in the nearby galaxies NGC 5253 and IC 4182 appear to give significantly smaller values in the range of $50 \lesssim H_o$ (km s$^{-1}$ Mpc$^{-1}$) $\lesssim 75$; the exact value depending on the adopted slope of the maximum magnitude versus rate of decline relation for SNe Ia and on the (still slightly controversial) light curves of SN 1937C and 1972E. A more fundamental problem is that recent observations by Hamuy et al. (1995) suggest that SNe Ia in galaxies with particularly young stellar populations, such as NGC 5253 and IC 4182, might have SNe Ia with above-average luminosities. If such a bias exists, it will lead to an under-estimate of $H_o$.

It is tentatively concluded that the best value for the age of the Universe is probably $t_o = 18 \pm 1 = 19$ Gyr, in which it is assumed that $\sim 1$ Gyr elapsed between the start of the Big Bang and the formation of the oldest globular clusters. Such an age clearly conflicts with that derived from $H_o$ in standard cosmological models. A similar conclusion has recently been drawn by Wambsganss et al. (1995) from a comparison of predictions of the frequency of lensed quasar separations with observation.

Turner et al. (1992) have argued that the local value of $H_o$ might differ significantly from its globular value. However, it appears unlikely that the apparent conflict between $H_o$ and $t_o$ can be resolved in this way because modern determinations of the Hubble parameter are based on observations of large volumes of space. From observations by Hoessel et al. (1980), van den Bergh (1992) finds that $H_o$(global) = $(0.94 \pm 0.07)$ $H_o$(local), in which $H_o$(local) refers to the region with redshift $V_o < 10000$ km s$^{-1}$. More recently, Lauer & Postman (1992) have used a full-sky sample of Abell clusters with $V_o < 15000$ km s$^{-1}$ to study variations in the Hubble parameter. These authors find that $H_o$ measured globally is constant to $\pm 7\%$ for $3000 < V_o$ (km s$^{-1}$) $< 15000$. These results show that there is presently no persuasive evidence for regional variations in $H_o$.



It is a pleasure to thank many of my colleagues for preprints and helpful discussions.

Particular thanks go to Ed Ajhar, Brian Chaboyer, Andy Gould, Peter Höflich, Andy Layden,

Dave Laney, Bev Oke and Matt Wood.

### *ERROR BUDGETS FOR CEPHEID DISTANCES*

| NGC 4571 | $\epsilon$ (mag) |
|---|---|
| LMC modulus | 0.10 |
| P - $L_R$ zero point | 0.04 |
| P - $L_R$ slope | 0.07 |
| $\sigma / \sqrt{n}$ | 0.12 |
| ptm. zero point | 0.02 |
| $\epsilon$ ($A_R$) | 0.00: (assumed) |
| | ______ |
| Total | 0.18 |

| M 100 | $\epsilon$ (mag) |
|---|---|
| LMC modulus | 0.10 |
| P - $L_I$ zero point | 0.03 |
| P - $L_I$ slope | 0.04 |
| $\sigma / \sqrt{n}$ | 0.05 |
| ptm. zero point | 0.04: |
| $\epsilon$ ($A_I$) | 0.08 |
| | ______ |
| Total | 0.15 |

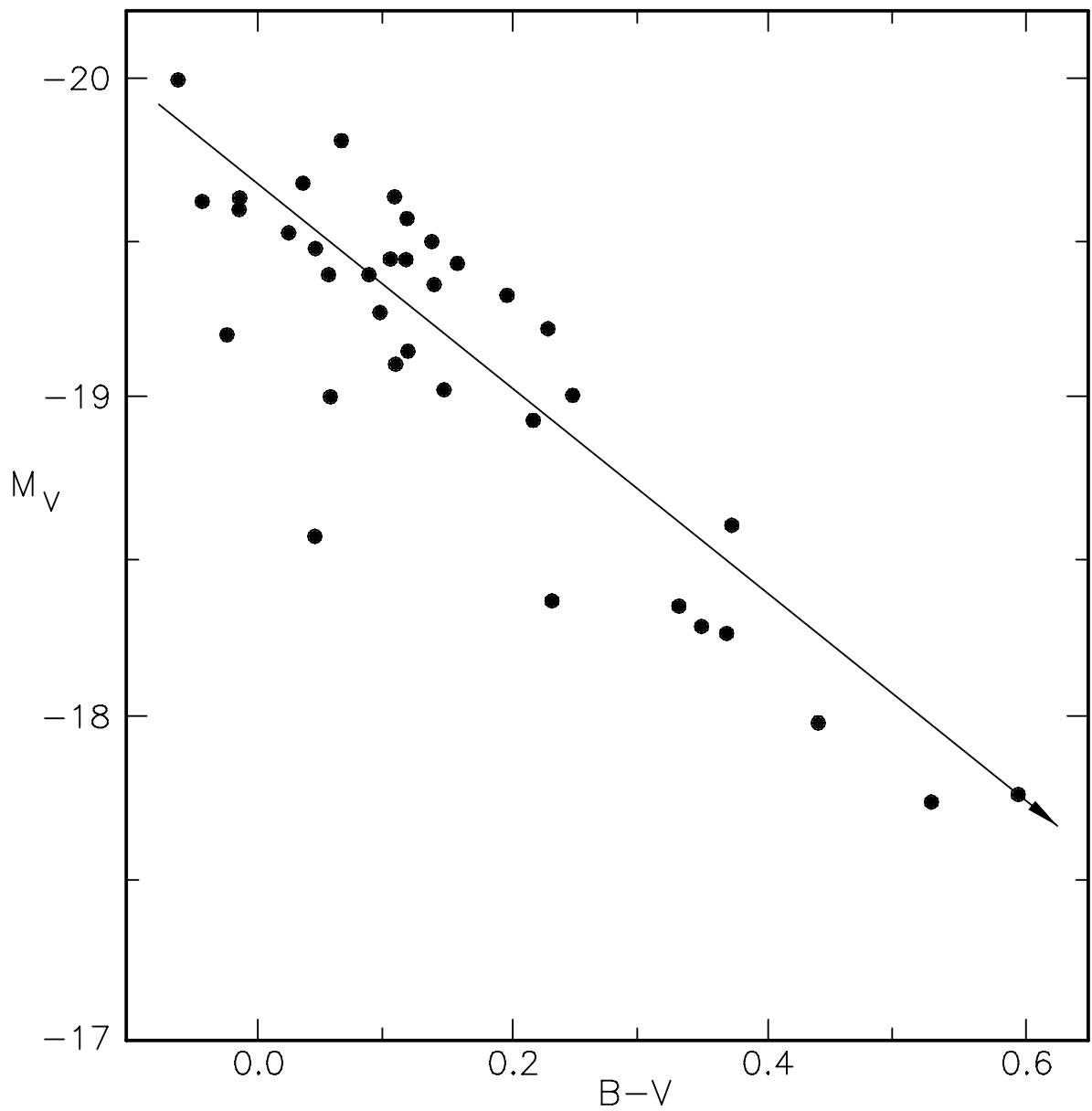